\documentclass[lettersize,journal,references]{IEEEtran}
\usepackage{amsmath,amsfonts}
\usepackage{algorithmic}
\usepackage{algorithm}
\usepackage{array}
\usepackage[caption=false,font=normalsize,labelfont=sf,textfont=sf]{subfig}
\usepackage{textcomp}
\usepackage{stfloats}
\usepackage{url}
\usepackage{verbatim}
\usepackage{graphicx}
\usepackage{subcaption}
\usepackage{multirow}
\usepackage{stfloats}
\usepackage{cite}
\begin{document}

\title{A Survey: RTE Solutions for Underwater Optical Communications}

\author{YunLong Li, Xiang Yi,~\IEEEmembership{Member,~IEEE,} ZhuoQi Chen, Peng Yue*,~\IEEEmembership{Member,~IEEE,}

\IEEEauthorblockA{State Key Laboratory of Integrated Services Networks, Xidian University, Xi'an, Shaanxi, 710071 China}

\IEEEauthorblockA{\IEEEauthorrefmark{1} Email: pengy@xidian.edu.cn}
}

\markboth{Journal of \LaTeX\ Class Files,~Vol.~14, No.~8, August~2021}%
{Shell \MakeLowercase{\textit{et al.}}: A Sample Article Using IEEEtran.cls for IEEE Journals}

\maketitle

\begin{abstract}
The Radiative Transfer Equation (RTE) is essential for solving the spatial distribution of light energy. It plays a crucial role in the link budget analysis of Underwater Wireless Optical Communication (UWOC). However, due to its complex integro-differential form, obtaining an exact solution is extremely challenging. This paper provides a systematic overview and comparison of key RTE solution strategies in case of UWOC scenario—including the Monte Carlo Method (MCM), Beer–Lambert Method (BLM), Beam Spread Function (BSF), Finite Element Method (FEM), and others—and analyzes how each approach balances accuracy, computational efficiency, and ease of implementation. Results show that MCM, though computationally intensive, to best match the three-dimensional spatial configuration of practical UWOC systems. BLM, while simplest, loses accuracy in turbid conditions. BSF partially corrects for scattering but yields only modest gains over BLM, and FEM struggles at longer ranges due to discretization. These findings help guide method selection for reliable estimation UWOC's system power budget.
\end{abstract}


\section{Introduction}
\IEEEPARstart{U}{nderwater} Wireless Optical Communication (UWOC) is a highly efficient data transmission technology with widespread applications in underwater environments, including ocean exploration, environmental monitoring, and intelligent underwater robotics \cite{baiden2007high,jing2023fast,chen2020vision,7593257}. Compared to traditional underwater acoustic communication, UWOC offers significantly higher bandwidth and lower latency. However, UWOC also faces unique challenges: the propagation of light in underwater environments is influenced by scattering, absorption, and multipath effects. These factors result in signal attenuation and distortion, limiting the communication distance and reliability of UWOC systems. Consequently, an accurate model for light propagation is critical for the design and analysis of UWOC. The Radiative Transfer Equation (RTE) is widely regarded as a fundamental mathematical tool for describing the spatial distribution of light energy in a random medium based on the principle of energy conservation \cite{chandrasekhar2013radiative,zaneveld1995light}.

RTE is crucial in estimating the spatial irradiance distribution in random scattering media. Unfortunately, due to its complex mathematical form,  exact analytic solution are hard to obtain \cite{arnon2012advanced,li2015use}. For many of these problems, the researcher must resort to numerical solutions\cite{xu2004electric,he2022analysis,ren2023simulation}. One of the most used solutions is MC photon tracing method (MCM). It tries to mimic the single photon's absorption and scattering events, and obtaining the statistical irradiance with tracing a large number of photons \cite{lerner1982monte,wang1995mcml}. It avoid to solve RTE directly, and therefore bypassing the mathematical difficulty. Hanson and Radic \cite{hanson2008high} compared the Monte Carlo simulation results with laboratory experimental results, demonstrating good agreement between the two. Leathers et al. \cite{leathers2004monte} provided a practical guide for generating Monte Carlo computer simulations for typical  applications. Generally, many researchers have also utilized the MCM to evaluate the path loss and channel impulse response (CIR) of underwater optical channels. \cite{6685978,xu2022improvement}. However, the MCM entails lengthy simulation times for long-distance scenarios and is susceptible to statistical errors.

Except for the widely used MCM, there still existed some numerical solutions. All of them tries to simplify the RTE with appropriate assumptions. As a numerical solution method for the 3D RTE, the Discrete Ordinates Method (DOM) discretizes the continuous directional and spatial domains, replacing the original integrals with a finite set of discrete directions and points. Studies related to DOM can be found in \cite{machida2022three} and \cite{ishimaru1978wave}, While DOM has certain advantages in handling multi-angle scattering, its uniform directional sampling often leads to over-sampling or under-sampling in scenarios where the scattering phase function is spatially uneven. Moreover, when the scattering phase function exhibits sharp peaks or the inherent optical properties (IOPs) vary significantly with depth, DOM can still face computational challenges. Building on this, the Finite Element Method (FEM) discretizes the angular space and reduces the 3D RTE to 2D RTE for efficient solving. Gao et al. \cite{gao2009fast} first suggested FEM as more suitable for strong forward scattering and employed uniform discretization in angular space. Later, Li et al. \cite{li2015use} proposed a non-uniform angular discretization method tailored to the actual distribution of the scattering phase function, significantly reducing computational costs. Illi et al. \cite{illi2019improved} further proposed a seven-point integration rule to optimize angular discretization and introduced an improved finite difference scheme, achieving simulation results for long-distance propagation.

On the other hand, inspired by advancements in image transmission, researchers have convolved the derived Point Spread Function (PSF) \cite{wells1969loss} with the light source to obtain the Beam Spread Function (BSF) \cite{cochenour2008characterization}, which simplifying the radiative transfer process to an analysis of beam property variations. Meanwhile, Xu \cite{xu2015analytical} further reduced the 3D RTE to two dimensions using the azimuthal single-scattering approximation and derived an analytical BSF solution. Yang et al. \cite{yang2024ulmc} applied this BSF model to analyze outage probability and average bit error rate (ABER) in ACO-OFDM systems. Notably, both FEM and BSF are two-dimensional solution methods, inherently restricting the  field of view (FOV) of receiver to \(180^\circ\). This represents a fundamental limitation of such approaches. 

At the 1D approximation level, the Invariant Embedding Method (IEM) is based on the principle of energy conservation and describes the spatial distribution of light energy in random media. However, there has been no direct application of the IEM to analyze the performance of UWOC channels. The Beer–Lambert Method (BLM), as the simplest 1D approximation, considers only single-pass attenuation in uniform media and does not account for multiple scattering or angular distribution effects. It is typically regarded as the theoretical bound when validating results of other complex methods \cite{mobley1994light}.

Although each method has its unique strengths, their performance and suitability vary significantly depending on the application context. This paper aims to conduct a systematic comparison of all methods, providing a detailed discussion of the advantages and limitations of each, comparing the performance of the MCM, BLM, BSF, and FEM based on simulation results, this study aims to provide valuable insights for researchers in the field of underwater optical communication. The ultimate goal is to guide the selection of the most appropriate computational method for specific application scenarios and to highlight potential directions for future technological advancements.

\section{Overview of Modeling Methods}
The scalar irradiance is determined by the radiance $L(r, \theta, \phi )$; 
$L$ is a function of five independent variables and is the solution of the 3D RTE \cite{zaneveld1995light}:

\begin{multline}
\frac{dL(r, \theta, \phi)}{dr} = 
- (a + b) L(r, \theta, \phi) \\
+ \int_{0}^{2\pi} \int_{0}^{\pi} 
L(r, \theta', \phi') \beta(r, \theta', \phi' \rightarrow \theta, \phi) 
\sin\theta' \, d\theta' \, d\phi' \\
+ S(r, \theta, \phi), \quad (\mathrm{W\, m^{-3}\, sr^{-1}})
\end{multline}
Here, $L(r, \theta, \phi)$ represents the radiance, which is a function of spatial coordinates $(r)$ and directional angles $(\theta, \phi)$. The $a$ is absorption coefficient and $b$ is scattering coefficient. The phase function $\beta(r, \theta', \phi' \to \theta, \phi)$ characterizes the probability of light scattering from direction $(\theta', \phi')$ to $(\theta, \phi)$. The source term $S(r, \theta, \phi)$ accounts for external or internal radiation sources, while the geometric factor $\sin\theta'$ accounts for angular projection effects. The radiance is measured in units of $\mathrm{W\, m^{-3}\, sr^{-1}}$, representing power per unit volume per unit solid angle. 

The solution of the 3D RTE is highly complex, particularly in non-uniform media. To address this challenge, the MCM offers an effective approach for solving the 3D RTE. MCM operates by simulating the propagation paths of a large number of photons and statistically estimating their contributions to the radiative intensity, thereby approximating the photon distribution. Meanwhile,  is a numerical approach that approximates the radiation distribution by discretizing the angular domain, dividing the spatial domain into grids, and solving the 3D RTE for each discrete direction individually.

These include approximations based on azimuthal single scattering, discretization of angular directions, and spatial domain discretization, which reduce the 3D RTE to 2D RTE. Building on this simplification, methods such as the BSF and FEM have been developed. These approaches provide either analytical or numerical solutions to efficiently solve the 2D RTE:

\begin{align}
\cos(\theta) \frac{\partial L(x, y, \theta)}{\partial x} + \sin(\theta) \frac{\partial L(x, y, \theta)}{\partial y} = - (a + b) L(x, y, \theta) \nonumber \\
+ b\int_{0}^{2\pi} P(\theta' \to \theta) L(x, y, \theta') d\theta' + S(x, y, \theta). 
\label{2D}
\end{align}
Furthermore, Mobley et al. \cite{mobley1994light} assumed that the light field is uniformly distributed within the horizontal plane and varies only with depth \( z \), resulting in the RTE for a 1D scenario:
\begin{multline}
\mu \frac{dL(z; \hat{\xi})}{dz} = - c(z)L(z; \hat{\xi}) 
+ \int_{\mathbb{H}} L(z; \hat{\xi}') \beta(z; \hat{\xi}' \to \hat{\xi}) d\Omega(\hat{\xi}') \\
+ S(z; \hat{\xi}),
\end{multline}

using the IEM to solve this equation, the Hydrolight model has demonstrated the accuracy of this approach. If there is no multiple-scattering, the RTE reduces to just:
\begin{equation}
\frac{dL(r, \theta, \phi)}{dr} = -c(r)L(r, \theta, \phi) .
\end{equation}
In source-free water, the solution is a simple exponential decay of the initial radiance with distance $L(r) = L(0)e^{-cr}$, this result is known as Beer-lambert's Law.

In Table \ref{tab:Various Methods}, we compare the advantages and limitations of these methods, and below, we provide a brief description of the methods to be compared in this paper.

\renewcommand{\arraystretch}{1.5} 
\begin{table*}[htbp]
\centering
\caption{Advantages and Limitations of Various Methods for Solving the RTE}
\label{tab:Various Methods}
\begin{tabular}{|>{\centering\arraybackslash}m{1.5cm}|>{\centering\arraybackslash}m{1.5cm}|>{\centering\arraybackslash}m{6.5cm}|>{\centering\arraybackslash}m{6.5cm}|}
\hline
\textbf{Dimension} & \textbf{Method} & \textbf{Advantages} & \textbf{Limitations} \\
\hline
\multirow{2}{*}{\centering \vspace*{-6em} 3D} & MCM & 
\begin{itemize}
\item completely general; can solve time-dependent and 3-D problems with arbitrary geometry
\item Provides clear statistical characteristics, making error estimation and uncertainty analysis straightforward
\item Widely applicable, capable of solving problems ranging from simple to highly complex scenarios
\end{itemize} &
\begin{itemize}
\item Computationally expensive, especially for large optical depths or intricate environmental conditions
\item Slow convergence, requiring a large number of photon traces to achieve stable solutions
\item Results depend on initial parameters and model assumptions, which can significantly affect outcomes
\end{itemize} \\

\cline{2-4}
& DOM &
\begin{itemize}
\item computed radiances do not have statistical errors
\item Easy to implement and understand, enabling quicker initial problem-solving approaches
\end{itemize} &
\begin{itemize}
\item does not handle highly peaked scattering phase functions well
\item Difficulty in managing boundary conditions, making it challenging for complex geometrical configurations
\end{itemize} \\
\hline

\multirow{2}{*}{\centering \vspace*{-7em} 2D} & BSF &
\begin{itemize}
\item Suitable for broadband signals
\item Simple computational approach that intuitively describes beam spreading in the medium
\end{itemize} &
\begin{itemize}
\item Limited accuracy
\item Dependence on simplified assumptions
\item Unsuitable for dynamic environments
\end{itemize} \\

\cline{2-4}
& FEM &
\begin{itemize}
\item High-accuracy solutions that capture detailed internal field distributions in complex media
\item Highly adaptable, allowing for flexible mesh refinement and sophisticated boundary condition handling
\item Supported by a solid theoretical foundation, facilitating integration with other numerical techniques
\end{itemize} &
\begin{itemize}
\item High computational complexity
\item Sensitivity to mesh quality, where poorly chosen discretization can degrade solution accuracy
\end{itemize} \\
\hline
\multirow{2}{*}{\centering \vspace*{-3em} 1D} & IEM &
\begin{itemize}
\item Efficient computation, providing rapid solutions for layered media scenarios
\item Ideal for analysis of hierarchically structured media
\item Includes all orders of multiple scattering
\end{itemize} &
\begin{itemize}
\item can solve only 1-D problems (the one dimension being the depth in optical oceanography)
\item difficult to program
\item Limited applicability, not easily extended to more complex geometrical configurations
\end{itemize} \\
\cline{2-4}
& BLM &
\begin{itemize}
\item Simple analytical expression
\item Provide a theoretical lower bound
\end{itemize} &
\begin{itemize}
\item Neglects multiple scattering
\item No angular distribution
\item No temporal dynamics
\end{itemize} \\

\hline
\end{tabular}
\end{table*}

\begin{enumerate}
\item{The MCM employs stochastic sampling to simulate photon interactions. By tracing the distribution of photons, it approximates solutions to the RTE, making it particularly effective for analyzing intricate scattering conditions and stratified water layers. Commonly applied to evaluate optical signal attenuation and angular spreading in underwater communication channels, the MC approach can yield accurate statistical characterizations—such as path loss as a function of distance—when conducted over a large number of trials. However, in medium- to long-distance scenarios, the MCM may exhibit slow convergence, with improving accuracy often requiring substantial computational resources and execution time. Moreover, limited computational resources can easily lead to statistical errors.}

\item{The BLM is a fundamental approach that models light attenuation along a LOS path in a homogeneous medium, accounting for absorption and scattering effects. It provides a straightforward analytical framework for estimating the attenuation of light intensity without considering multiple scattering or angular distribution changes. As one of the simplest techniques, the Beer-Lambert method offers a theoretical bound, making it a valuable tool for verifying the reasonableness of results from more complex models. However, its simplicity also imposes significant limitations. It is best suited for idealized, uniform media and cannot accommodate scenarios involving non-homogeneous environments, multiple scattering events, or dynamic angular variations. This limits the BLM to providing theoretical reference only in scenarios where the water is not highly turbid.}

\item{The BSF method streamlines radiative transfer analysis by concentrating on how the profile of a beam evolves, deforms, and disperses while traveling through water. Rather than tackling the full intricacy of light transmission, it spotlights how beam shape and intensity distribution shift under the influence of scattering and absorption. Often employed for far-field optical modeling, the BSF technique enables rapid channel characterization and provides an intuitive understanding of spatial beam patterns. Its simplified representation does, however, have limitations in highly intricate or strongly multipath scenarios. Analyses typically rely on basic assumptions—such as approximating beams as Gaussian—which may not fully capture the complexity inherent in actual oceanic conditions. These limitations become more pronounced as the turbidity of the water increases.}

\item{The FEM divides the spatial and angular domains into discrete elements, transforming a continuous RTE-based problem into a finite system of equations. This approach supports high-fidelity simulations of light transmission in non-uniform water masses and can adeptly accommodate intricate boundary configurations. For example, in the case of Eq.(\ref{2D}), the FEM subdivides the XoY plane into small grids, enabling it to account for underwater medium inhomogeneity and complex boundary conditions. Simple basis functions are applied within each grid element, and the local integral results from all elements are assembled to transform the partial differential equation into a discrete integral equation. This is further formulated into a sparse nonlinear algebraic system, which is then solved. . However, the quality of the mesh significantly influences both
accuracy and efficiency, as poorly constructed meshes
can induce errors and slow execution. Therefore, FEM
can only provide rapid and effective solutions over short
to medium distances}

\end{enumerate}

\section{Versus the Result}
In this section, we investigate the applicability of the MCM, BSF, FEM and BLM across different propagation distances and aquatic environments. Using the MCM outlined in \cite{pharr2023physically}, the experimental parameters are configured as follows: the transmitter emits \(10^8\) photons, the FOV of the receiver is set to \(180^\circ\) and the aperture diameter to 0.1 m to ensure a fair comparison with the BSF and FEM models. Additionally, the IOPs for various water types, as detailed in Table~\ref{tab:iops}, are incorporated into the simulations. The primary aim of this study is to assess whether the approximate solutions of the RTE remain valid across different propagation distances and water conditions. To improve computational efficiency while maintaining comparative accuracy, all simulations use the HG scattering phase function as a simplified model, with the anisotropy factor g set to 0.924..

\begin{table}[h]
\centering
\caption{IOPs for Different Water Types}
\label{tab:iops}
\begin{tabular}{|l|c|c|c|}
\hline
\textbf{Water Types} & \(a(\lambda)\) (\(m^{-1}\)) & \(b(\lambda)\) (\(m^{-1}\)) & \(c(\lambda)\) (\(m^{-1}\)) \\
\hline
Pure sea water       & 0.053                        & 0.003                        & 0.056                        \\
\hline
Clear ocean water    & 0.114                        & 0.037                        & 0.151                        \\
\hline
Coastal ocean water  & 0.179                        & 0.219                        & 0.398                        \\
\hline
Turbid harbor water  & 0.295                        & 1.875                        & 2.170                        \\
\hline
\end{tabular}
\end{table}

\begin{table*}[b] 
    \centering
    \begin{tabular}{cc} 
        \begin{minipage}[t]{0.48\textwidth}
            \centering
            \includegraphics[width=\textwidth]{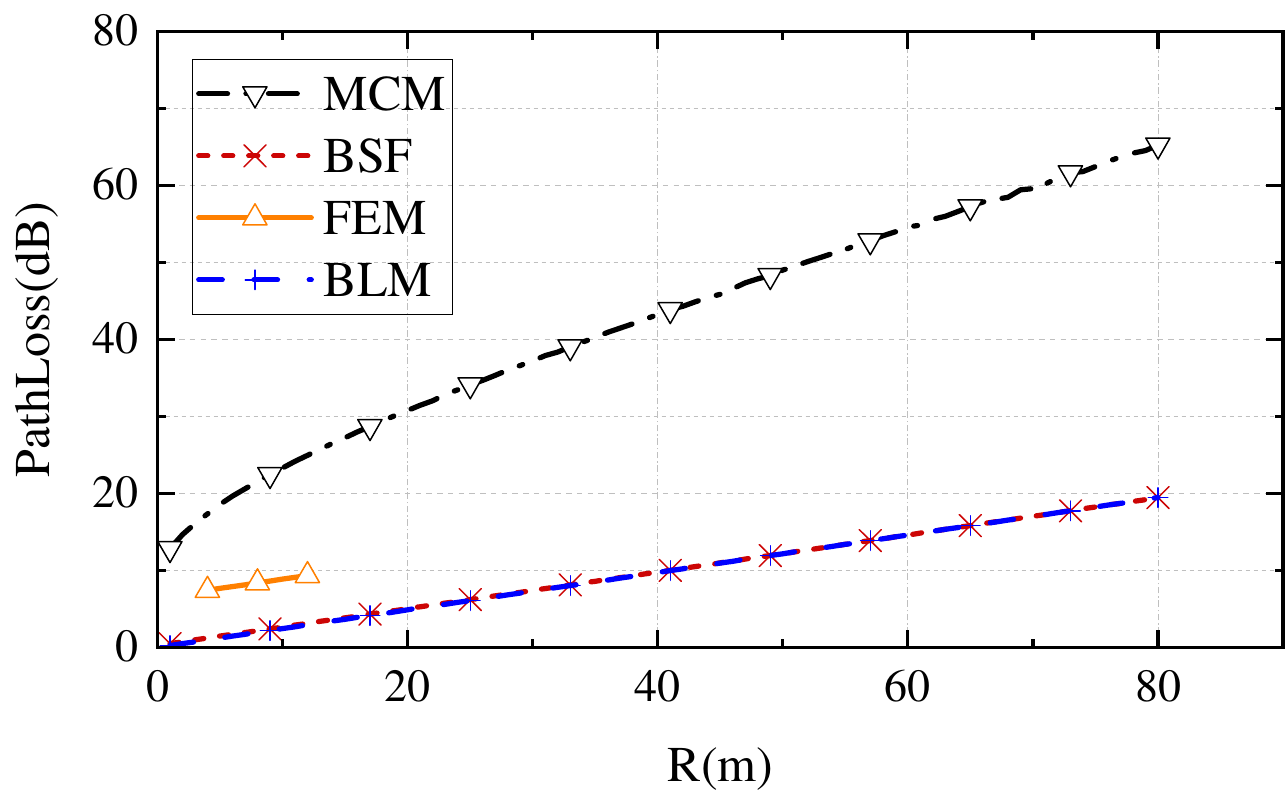}
            \captionof{figure}{Pure sea water}
            \label{fig:image1}
        \end{minipage} &
        \begin{minipage}[t]{0.48\textwidth}
            \centering
            \includegraphics[width=\textwidth]{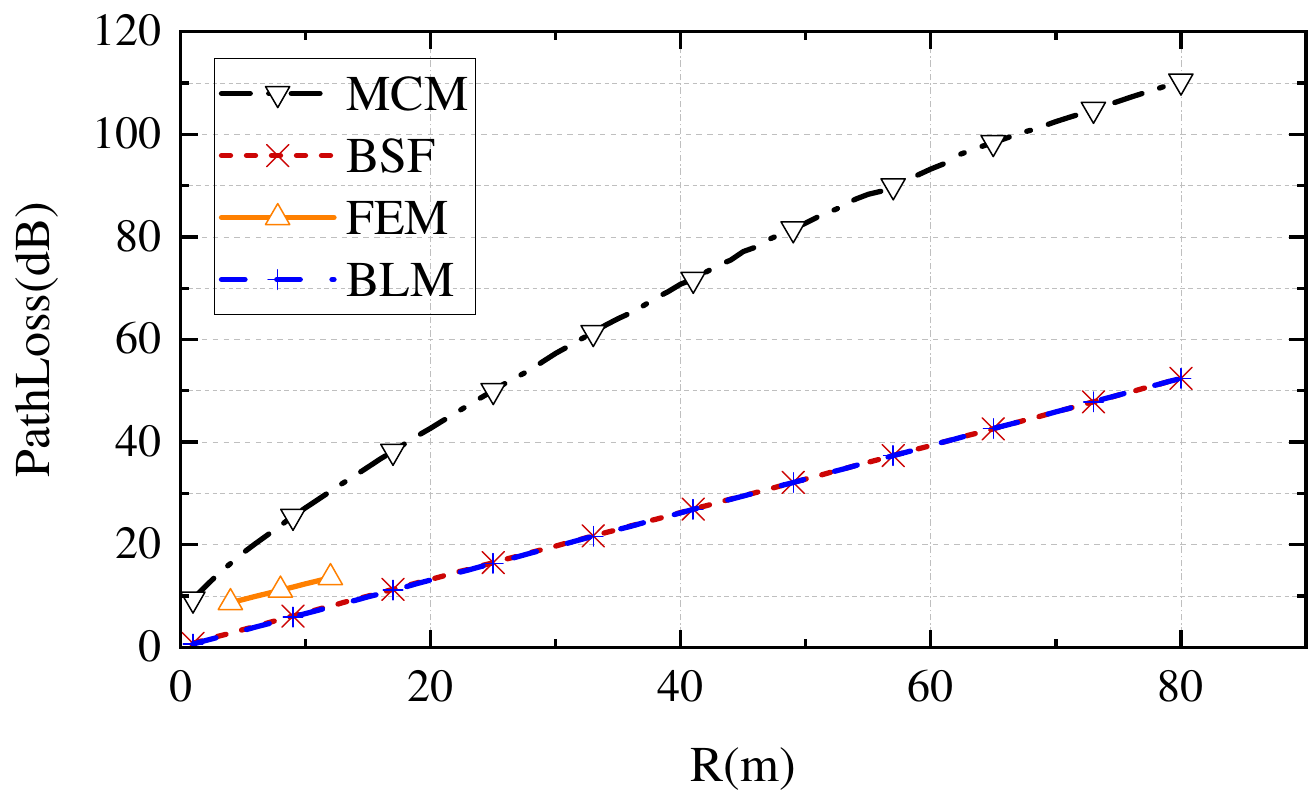}
            \captionof{figure}{Clear ocean water}
            \label{fig:image2}
        \end{minipage} \\
        \begin{minipage}[t]{0.48\textwidth}
            \centering
            \includegraphics[width=\textwidth]{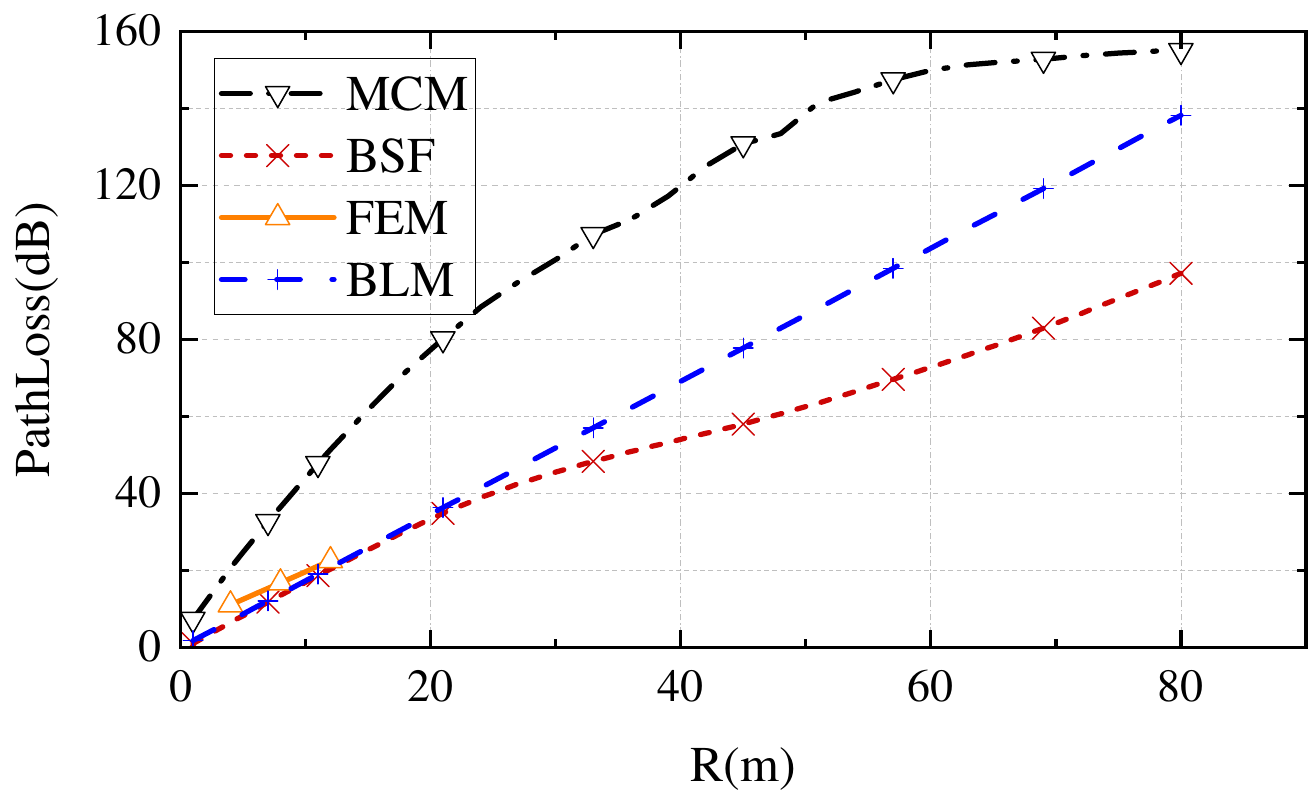}
            \captionof{figure}{Coastal ocean water}
            \label{fig:image3}
        \end{minipage} &
        \begin{minipage}[t]{0.48\textwidth}
            \centering
            \includegraphics[width=\textwidth]{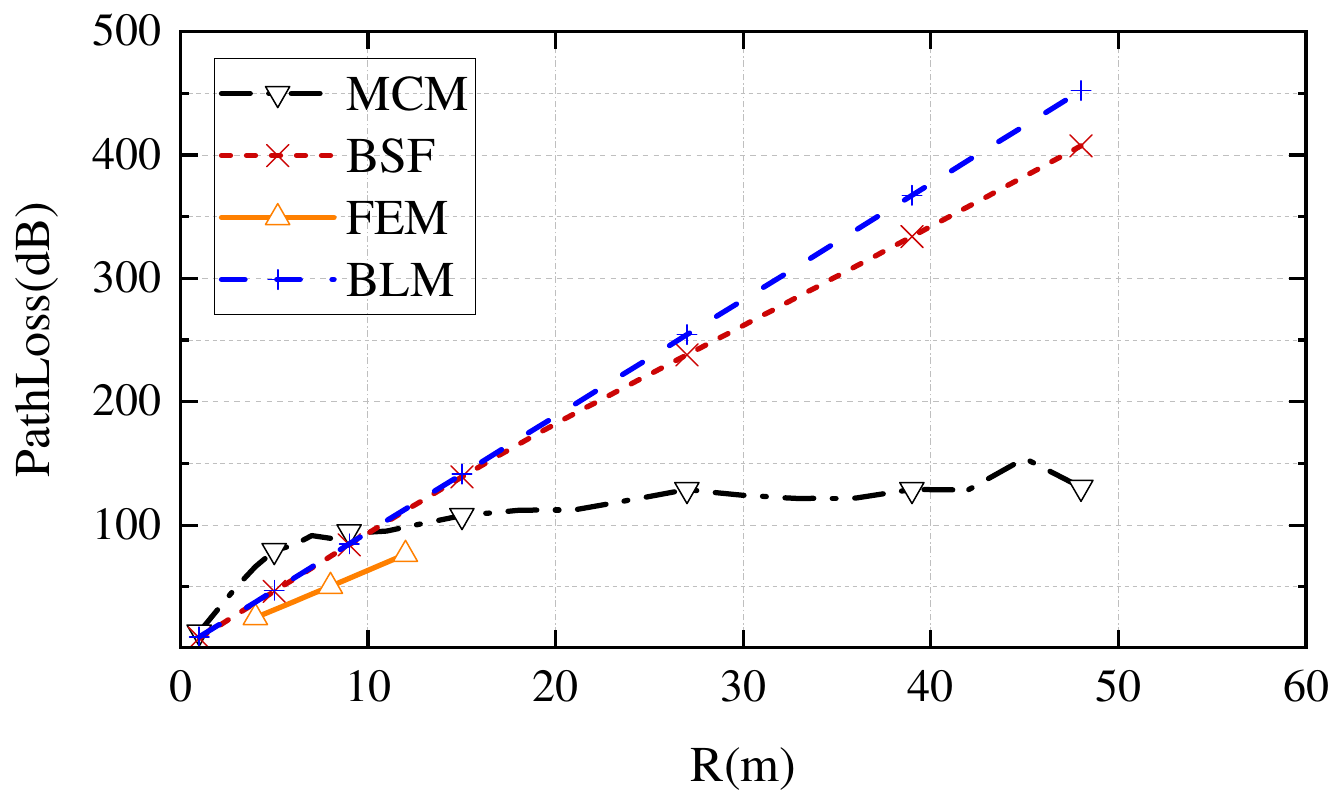}
            \captionof{figure}{Turbid harbor water}
            \label{fig:image4}
        \end{minipage} \\
    \end{tabular}
    \label{fig:all_images}
\end{table*}

Figure \ref{fig:image1}-\ref{fig:image4} presents a comparison of the results obtained using MCM, BSF, FEM, and BLM under four different water conditions. The MCM is widely regarded as the most accurate numerical simulation technique because it comprehensively models the stochastic processes of photon multiple scattering and absorption in water. This makes it highly reliable across various water quality conditions. In contrast, the BLM is a fundamental exponential attenuation model, suitable for scenarios where scattering is negligible or only simplified single-scattering effects are considered. However, as water quality transitions from clear to turbid, the effects of multiple scattering become more pronounced, leading to significant deviations in BLM predictions (typically resulting in overestimation or underestimation). The BSF method builds upon BLM by incorporating corrections for small-angle scattering and transmission path adjustments. However, as the results show, these scattering corrections offer negligible improvement in clear and pure water conditions. As water turbidity increases, the results of BSF also exhibit significant deviations. The FEM method accounts for partial scattering effects, demonstrates good reliability across four types of water bodies at short distances. But struggles with increasing propagation distances and cumulative scattering. As a result, its accuracy deteriorates rapidly due to discretization and grid approximation errors.

As shown in Figures \ref{fig:image1} and \ref{fig:image2}, the MCM exhibits the highest and fastest-growing path loss curve in clear or pure water environments. This is because the MCM tracks numerous photon events, including multiple scattering and absorption within the water. The cumulative effect significantly reduces the number of effective photons reaching the receiver. In such conditions, where the overall attenuation coefficient of the water is relatively small, the exponential decay curves of BLM and BSF show minimal growth. FEM partially corrects for scattering effects, resulting in path loss values that lie between those of MCM and BLM. However, due to limitations in grid discretization and boundary conditions, FEM can only simulate path loss effectively over limited distances. In coastal waters, as shown in Figure \ref{fig:image3}, scattering effects become more pronounced. MCM captures more multiple scattering events, leading to higher path loss. In contrast, BLM, which only accounts for simple absorption and single scattering, begins to exhibit significant errors due to its simplified assumptions. BSF partially corrects for scattering contributions, causing its results to diverge from those of BLM. FEM also shows higher path loss in these waters compared to clear waters but remains below the estimates provided by MCM. In the highly turbid water scenario depicted in Figure \ref{fig:image4}, the intense scattering exceeds the applicability of simple exponential decay models. As a result, BLM and BSF experience severe energy attenuation with increasing distance. FEM also shows a significant increase in path loss in such water conditions, but its values remain lower than those of the other methods. Notably, the MCM curve shows unexpectedly lower path loss at long distances. This may be attributed to the extreme multiple scattering and absorption effects causing a large proportion of photons to be lost prematurely, leading to insufficient photon counts at the receiver and statistical errors. Another potential reason is the assumed forward-scattering phase function, which may inadvertently allow a small portion of photons to retain a directional bias toward the receiver during random processes. These photons contribute disproportionately to the Monte Carlo results, outweighing photons that are scattered at large angles or fully absorbed, thereby lowering the overall macroscopic attenuation value.

\section{Conclusion}
In this study, we compared the performance of different models across various water environments, ranging from clear to highly turbid waters. The results highlight that each light propagation model has its strengths and limitations in terms of accuracy and applicability. In regions with relatively weak scattering effects, BLM and BSF provide rapid and reasonably accurate attenuation predictions, making them suitable for large-scale preliminary assessments. For scenarios requiring high-fidelity simulations of local scattering and boundary conditions, FEM performs better over short distances. MCM, while computationally intensive and time-consuming, excels in capturing multiple scattering effects and complex water structures, making it nearly indispensable in high-turbidity environments. Therefore, selecting an appropriate solution method requires a comprehensive consideration of water turbidity, system range, computational cost, and the desired accuracy. A balanced choice of model can ensure precise evaluations of underwater communication performance and enable optimal system design.


%

\bibliographystyle{IEEEtran}

\bibliography{references}

\newpage

\end{document}